\documentstyle[prd,aps,floats,epsf]{revtex}

\def\bea{\begin{eqnarray}}
\def\eea{\end{eqnarray}}

\def\tQ{\tilde Q}

\newcommand{\beq}{\begin{equation}}
\newcommand{\eeq}{\end{equation}}

\newcommand{\pbs}{{\varphi_{b}^{60}}}

\begin{document}
\preprint{BROWN-HET-1347, hep-ph/yymmddd}
\draft 

%
% Remove this and closure after abstract, plus preprint number,
% in electronic submission
%
 
%
\renewcommand{\topfraction}{0.99}
\renewcommand{\bottomfraction}{0.99}
\twocolumn[\hsize\textwidth\columnwidth\hsize\csname 
@twocolumnfalse\endcsname
  
\title
{\Large On the Initial Conditions for Brane Inflation}
       
\author{R. Brandenberger, G. Geshnizjani and S. Watson} 
\address{~\\Department of Physics, Brown University, 
Providence, RI 02912, USA}
\date{\today} 
\maketitle
\begin{abstract} 
String theory gives rise to various mechanisms to generate
primordial inflation, of which ``brane inflation'' is one of
the most widely considered. In this scenario, inflation takes
place while two branes are approaching each other, and the
modulus field representing the separation between the branes
plays the role of the inflaton field. We study the phase
space of initial conditions which can lead to a sufficiently
long period of cosmological inflation, and find that taking
into account the possibility of nonvanishing initial momentum
can significantly change the degree of fine tuning of the required initial
conditions.         
\end{abstract}

]

\vskip 0.4cm

\section{Introduction}

The paradigm of cosmological inflation \cite{Guth} has been
spectacularly successful as a theory of the very early Universe.
Not only does having a very early period of cosmological inflation 
solve some of the mysteries of Standard Big Bang cosmology such as
the horizon and the flatness problems, but it also gives rise to a
mechanism which generates the primordial fluctuations required to
explain today's cosmological large-scale structure and the observed
anisotropies in the cosmic microwave background (CMB). Simple models
of inflation rather generically predict a scale-invariant spectrum
of adiabatic cosmological fluctuations, a prediction which has
recently been verified with significant accuracy by CMB anisotropy
experiments \cite{Boomerang,Maxima,Dasi,Archaeops,WMAP}.

However, at the present time the paradigm of cosmological inflation
is lacking an underlying theory. The inflaton, the scalar field which
is postulated to generate the quasi-exponential expansion of the Universe,
cannot be part of the standard particle physics model, nor does it
fit easily into pure field theory extensions of the standard model
(see e.g. \cite{RHBrev} for a recent review of progress and problems
in inflationary cosmology). In particular, it is not easy to justify
a pure field theory based model of inflation in which the potential
of the inflaton is sufficiently flat in order not to produce too large
an amplitude of the fluctuation spectrum. Moreover, in many simple scalar
field toy models of inflation,
the overall amplitude of the spectrum of fluctuations points to a 
scale of inflation which is close to the scale of
particle physics unification, and thus close to the scale where new
fundamental physics, e.g. string theory, will become important. Thus,
it is natural to look for realizations of inflation in the context of
string theory.

One of the key observations which makes it promising to consider
string theory as the source of cosmological inflation is the fact
that string theory contains many scalar fields (the moduli fields)
which are massless
in the absence of non-perturbative effects and in the absence of
supersymmetry breaking. Hence, it is natural to look for ways of
obtaining inflation from moduli fields. Recent developments in
string theory have led to new possibilities. In particular, based
on the observation that string theory contains p-branes as
fundamental degrees of freedom, and that the matter fields of the
Universe can be considered to be localized on these branes, the
brane world scenario has emerged in the past few years offering
a complete change in our view of the Universe \cite{brane}.

Within the context of brane world scenarios, there are new ways
to obtain inflation. For example, if the brane which represents
the space-time on which our matter fields are confined is moving
in a nontrivial bulk space-time, it is possible to obtain
inflation on the brane from the dynamics in the bulk (``mirage
inflation'' \cite{Kraus,Kehagias,Alexander}). Another possibility
is topological inflation on the brane \cite{Alexander2}. The
approach which was first suggested in \cite{Dvali} and has received most 
attention in the literature is ``brane inflation'', a scenario in
which inflation on the brane is generated while two branes are
approaching each other. In this scenario, the separation of the
branes plays the role of the inflaton. 

The basic idea of brane inflation is the following \cite{Dvali}: consider
two parallel BPS branes (see e.g. \cite{Pol} for the string theory
background). When they sit on top of each other the
vacuum energy cancels out. In the cosmological setting, they
will start out relatively displaced in the extra
dimensions. There is then an attractive force between the branes
due to exchange of closed string modes, the stringy realization of
the gravitational attraction between the branes. If supersymmetry
is unbroken, the gravitational attraction is cancelled out by the
repulsion due to the Ramond-Ramond (RR) field. The separation between
the two branes acts as a scalar field in the effective field theory on
the brane. With two parallel BPS branes, the potential for this scalar
field is flat before supersymmetry breaking, and hence no cosmological
inflation can be induced. After supersymmetry breaking, only the
graviton remains massless, and hence there will be a net gravitional
attractive force between the branes. The induced
nonperturbative potential is flat and hence can yield a period of
cosmological inflation on the brane while the branes are widely separated.

{}Following the pioneering work of \cite{Dvali}, several concrete
realizations of the basic scenario were proposed which have the
advantage of allowing for controlled computations of the inter-brane
potential (see \cite{Quevedo} for a recent review). 
One realization \cite{Burgess1,Shafi} which will be 
analyzed in Section 3 is
based on a D4 brane-antibrane pair in Type IIA string theory
compactified on an internal torus $T^6$. The brane pair is 
parallel but separated in internal space. In this case, the
potential can be computed explicitly. It turns out that a sufficient
period of inflation can be obtained provided the branes are located
approximately at the antipodal points. This leads to a severe
initial condition constraint on the model. By fixing the branes to orientifold
fixed planes \cite{Burgess2} this initial condition constraint
can be somewhat alleviated. Another brane inflation
scenario \cite{Kyae,Juan,Tye,Zavala} 
is based on taking two D4 branes which again
are parallel except for forming an angle $\theta$ (taken to be
small) in one of the 
two-dimensional hyperplanes. In this case, the attractive potential is
once again computable in string perturbation theory, and the
resulting potential is suppressed by the small parameter $\theta$, thus
making the probability to obtain sufficient inflation larger. 
The brane inflation paradigm appears in fact to be rather robust.
Models based on the attractive force between branes of different
worldsheet dimensionalities were considered in 
\cite{Kallosh1,Kallosh2,Blumenhagen,Brodie}. 

A question which has not yet been addressed in the literature on
brane inflation is the issue of constraints on the phase space
of initial conditions for inflation which arise when one takes
into account the fact that in the context of cosmology the momenta
of the moduli fields which give inflation cannot be neglected
in the early Universe. For simple scalar field
toy models of inflation, the constraints on the phase space of
initial conditions taking into account the range of allowed
initial values not only for the inflaton field, but also for its
momentum, were studied in 
\cite{ABM,Goldwirth1,Goldwirth2,Kung1,Kung2,Goldwirth3,Goldwirth4}. 
As this work showed,
there is a big difference in the degree of tuning required in order
to obtain sufficient inflation between {\it small-field} inflation models
of the type of {\it new inflation} \cite{Albrecht,Linde82} and
{\it large-field} inflation models such 
as {\it chaotic inflation} \cite{Linde83} \footnote{See \cite{Dodelson}
for the classification of inflationary models into those of small-field
and large-field type.}.

For models of the type of {\it new inflation}, the work of 
\cite{Goldwirth1,Goldwirth2} showed that allowing for
nonvanishing intial field momenta may dramatically
reduce the phase space of initial conditions 
for which successful inflation results. This result is easy
to understand: in new inflation, the field must start off close to
an unstable fixed point, from which point it then slowly rolls to its
minimum. Unless the initial field momentum is finely tuned (as finely or even
more finely that the field value), slow rolling is never realized.
In contrast, in models of the type of {\it chaotic inflation}, most of
the energetically accessible field value space gives rise to a sufficiently
long period of slow roll inflation. Since the momentum redshifts much
faster than the potential energy, the initial kinetic energy does not
have to be small in order that the field will settle down close to the
slow-roll trajectory (see also \cite{Felder} for a recent extensive
phase space study of homogeneous inflationary cosmology) \footnote{Note
that the studies of \cite{Goldwirth1,Goldwirth2,Felder} mentioned
above were performed without taking inhomogeneities into account.
As studied in \cite{Goldwirth3} for new inflation and in 
\cite{Kung1,Kung2,Goldwirth3,Goldwirth4} 
for chaotic inflation, including spatial inhomogeneities
accentuates the difference between models like new inflation and those
like chaotic inflation. Inhomogeneities further reduce the measure
of initial conditions yielding new inflation, whereas the inhomogeneities
have sufficient time to redshift in chaotic inflation, letting the
zero mode of the field eventually drive successful inflation. The
property of chaotic inflation as an attractor in initial conditon
space \cite{Kung2} persists even when including linearized gravitational
fluctuations \cite{Feldman}.} Note that
{\it natural inflation} models \cite{natural}
fall into the former category (and thus
are not natural at all from the dynamical systems perspective), whereas
{\it hybrid inflation} \cite{hybrid} 
models are more like chaotic inflation in terms
of the initial condition aspect. 

In the following section, we review the constraints on the initial
conditons in phase space required for successful inflation and formulate
the argument in a form which is applicable to general models of inflation
driven by a single scalar field. This then allows us to apply (in Section 3)
the arguments to a couple of representative models of brane inflation. 
We find that in some models of brane inflation, in particular in
the case of branes at an angle,  there is dramatic 
reduction to the relative probability of obtaining successful
inflation when allowing the initial momentum to be general when compared
to what it obtained allowing only the initial field position to vary and
taking the initial momentum to vanish. In these models, the
sensitivity of brane inflation to initial conditions is worse than in
the case of chaotic inflation.

\section{Method}

There have been a large number of inflationary models proposed over the
years.  These models can be divided into two classes, {\it large
field} (e.g. chaotic inflation) and {\it small field} inflation (e.g. new
inflation).  In all these models slow roll inflation must last
long enough to solve the flatness, horizon, and monopole
problems. This requires \cite{Guth} that the Universe inflate 
during slow roll by at least ${\mathcal N}=60$ e-foldings. 
In the following, we develop
a general method for determining the constraints on the phase space
of initial data resulting from enforcing these requirements.
Specifically, we want to consider the allowed values for the
initial kinetic term that will still allow for adequate inflation.

We will first give some general considerations applicable to both
classes of inflationary models, and only later consider separately 
the cases of small field and large field models.
We will find that in the latter case one can achieve adequate
inflation quite naturally, while the former suffers from
fine-tuning of initial conditions if one wants successful inflation
and considers the full energetically allowed phase space of initial 
conditions for both the inflaton field and its momentum.

We will restrict our attention to homogeneous evolution of the
inflaton field. Small inhomogeneities could be included using
the methods of \cite{Feldman}.

\subsection{General Phase Space Considerations}

We want to find which regions of the phase space $(\varphi, \Pi)$ 
will lead to successful inflation (here, $\varphi$ is the inflaton field and
$\Pi \equiv \dot{\varphi}$ is its velocity). From the point of
view of early Universe cosmology it is unreasonable to postulate
that $\varphi$ starts at some distinguished value (as was initially
assumed in models of new inflation), nor is it justified to set
the initial momentum to zero. The full phase space of initial
conditions depends on the specific inflationary model being
considered. For example, in models in which inflation is to
commence immediately after the Planck time, a reasonable condition
is to bound the phase space by requiring that the total energy
density be smaller than the Planck density. Given initial
conditions anywhere in this allowed phase space, we ask what
subset of these initial conditions will lead to trajectories with
a sufficiently long period of slow roll inflation (inflation 
not necessarily starting immediately at the initial time).

The {\it slow roll} conditions for inflationary dynamics single
out a special trajectory $\Gamma$ in phase space. In order to have 
successful slow roll inflation, the phase
space trajectory of the inflaton field 
must, during some time interval, be sufficiently close to the portion
of the trajectory $\Gamma$ which yields a sufficient number of
e-foldings of inflation in order to solve the cosmological problems
mentioned at the beginning of this section. Our method (motivated by
the work of \cite{Goldwirth1,Goldwirth2}) 
is to follow these trajectories backwards in
time to uncover the entire phase space that leads to successful
inflation.

Let us begin by recalling the relevant equations of motion, which
are the field equation for the
inflaton and the Friedmann equation,
\begin{equation} \label{eom1}
\dot{\Pi}+3H\Pi+{V^{\prime}}(\varphi)=0,
\end{equation}
\begin{equation} \label{eom2}
H^{2}=\frac{8 \pi}{3 M_{p}^{2}} \Bigl( \frac{1}{2} \Pi^{2}
+V(\varphi) \Bigr).
\end{equation}

\begin{figure}
\epsfxsize=2.9 in \epsfbox{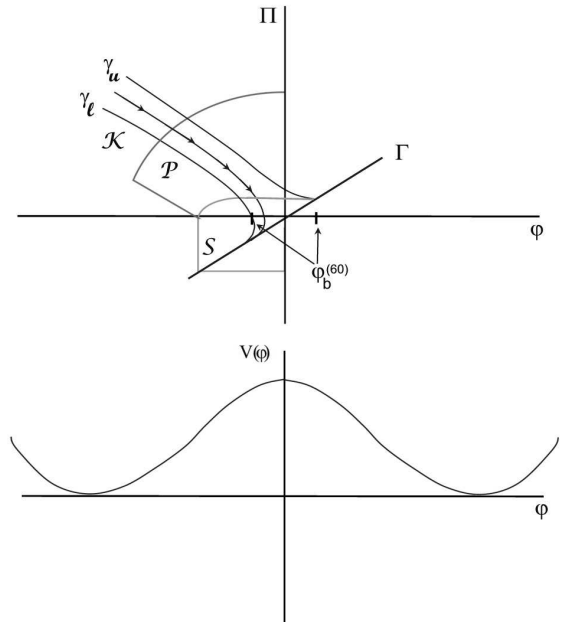}
\caption{Sketch of the potential (lower figure) and phase space in a
model of small field inflation. The region which leads to successful
slow roll inflation lies between the the lower and upper diagonal
curves $\gamma_l$ and $\gamma_u$ in the upper left quadrant. 
The symbols are defined in the text.}
\label{fig1}
\end{figure}

Given these equations we proceed by dividing the phase space into
regions $\mathcal{S}$, $\mathcal{P}$, and $\mathcal{K}$ as
indicated in Figures 1 and 2. The region $\mathcal{P}$
is the region where the potential energy dominates over the
kinetic energy, but the slow roll conditions (even in the
generalized sense discussed below) are not satisfied, and
the $\mathcal{K}$ is the region of phase space where the
kinetic energy dominates.

\begin{figure}
\epsfxsize=2.9 in \epsfbox{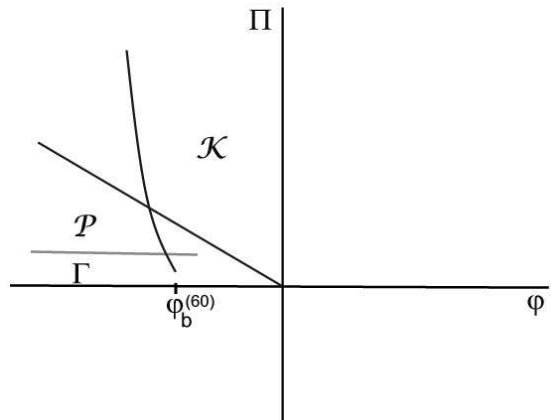}
\caption{Sketch of the phase space in a
model of large field inflation. The region which leads to successful
slow roll inflation lies to the left of the steep curve in 
the upper left quadrant.}
\label{fig2}
\end{figure}

Region $\mathcal{S}$ is
the region of {\it generalized slow roll}. To define this region,
let us remind the reader that the usual slow roll conditions
of inflationary dynamics state that
we can neglect the kinetic energy compared
to the potential in (\ref{eom2}) and $\dot{\Pi}$ compared to
$V^{\prime}$ in (\ref{eom1}). Under these assumptions, the equations
become
\begin{equation}\label{sra1}
3H\Pi+{V^{\prime}}(\varphi)=0,
\end{equation}
\begin{equation} \label{sra2}
H^{2}=\frac{8 \pi}{3 M_{p}^{2}} V(\varphi).
\end{equation}
The solution to these equations is a curve $\Gamma$ in phase space. 
As has recently been studied in detail in \cite{Felder}, in
an expanding Universe, this curve is an attractor curve. The
curve $\Gamma$ does not extend to all values of $\varphi$,
since the slow roll conditions break down.

There is a region ${\cal S}$ of phase space which forms a very narrow
strip about the curve $\Gamma$ and which extends from the curve to
the $\Pi = 0$ axis and slightly beyond, where the potential energy
density dominates over the kinetic energy density, but where in the
scalar field equation of motion neither the $V^{\prime}$ nor the
${\dot \Pi}$ terms are negligible (see Figure 1). A phase space
trajectory must enter region ${\cal S}$ if slow roll inflation is
to occur at all. 

Along the curve $\Gamma$ there is a distinguished value of $\varphi$
which we will denote as $\varphi_b^{60}$. If the ``beginning''
value $\varphi_b$ of $\varphi$ obeys 
\begin{equation} \label{srcond}
V(\varphi_b) \, > \, V(\varphi_b^{60}) \, ,
\end{equation} 
then the trajectory will
experience a sufficient number ${\mathcal N}= 60$ of e-foldings of inflation. 
By integrating the slow roll equations (\ref{sra1},\ref{sra2}), we find
that $\mathcal{N}$ is given by
\begin{equation} \label{e-fold}
{\cal{N}}=8\pi M^{-2}_{p} \int_{\varphi_{e}}^{\varphi_{b}}
\frac{V}{V^{\prime}} \; d\varphi \geq 60,
\end{equation}
where ${\varphi}_{e}$ is the value of the field where the slow roll
approximation breaks down.
For the case of small field inflation \footnote{Without loss of
generality we may assume for small field inflation
that $\varphi = 0$ is the local maximum of the potential near which
inflation occurs, and for simplicity we take the potential to
be symmetric about $\varphi = 0$. In the case of large
field inflation, we take $\varphi = 0$ to be the minimum
of the potential, and again we take the potential to be
symmetric about this field value.} the constraint (\ref{srcond}) 
is given by
$|\varphi_{b}| \leq \varphi_{b}^{60}$ and $|\varphi_{b}| \geq
\varphi_{b}^{60}$
gives the constraint for large field models (see Figure 2).

{}From the region $\mathcal{S}$ we trace the evolution
backwards to the region $\mathcal{P}$.  In this region the
potential still dominates the kinetic term, but instead of the ${\dot \Pi}$
term, it is the $V^{\prime}$ term in the scalar field equation of
motion which is negligible.
Thus, in this region (\ref{eom1}) and (\ref{eom2}) become,
\begin{equation} \label{P1}
\dot{\Pi}+3H\Pi=0,
\end{equation}
\begin{equation} \label{P2}
H^{2}=\frac{8 \pi}{3 M_{p}^{2}} V(\varphi).
\end{equation}

Equations (\ref{P1}) and (\ref{P2}) for a general form of the potential
can be combined into one equation,
\begin{equation}\label{generalp}
\Pi_{pk}-\Pi \, = \,
-\frac{\sqrt{24\pi}}{M_{p}}\int_{\varphi}^{\varphi_{pk}}\sqrt{V(\varphi)}d\varphi \, ,
\end{equation}
where $\varphi_{kp}$ and $\Pi_{kp}$ are the values at the boundary
between regions ${\mathcal P}$ and ${\mathcal K}$. If we assume
that $V(\varphi)$ is almost constant in region ${\mathcal P}$ the
above relation
simplifies to a linear relation between $\varphi$ and $\Pi$ in this region,
\begin{eqnarray}\label{region2eq2}
\varphi={\varphi}_{kp}+\frac{({\Pi}_{kp}-\Pi)M_{p}}{\sqrt{24\pi
V(\varphi)}},
\end{eqnarray}
The consistency of this approximation must be investigated for
each potential separately. The condition for the approximation to
be good can be written as (within a first order analysis):
\begin{equation}\label{condforV}
\Biggr| \frac {1}{4} \frac {V^{\prime}(\varphi)}{V(\varphi)}\delta\varphi \Biggr|
\ll 1 \, .
\end{equation}
As will be discussed later, this condition is not satisfied 
for all models of inflation.

Let us now denote by $\varphi_{b}$ and $\Pi_{b}$ the values of
the inflaton and its momentum when the trajectory enters the generalized
slow roll region $\mathcal{S}$. These values are related to
$\varphi_{kp}$ and $\Pi_{kp}$ via (\ref{region2eq2}), and thus
\begin{equation}\label{PSboundary1}
\varphi_{b}=\varphi_{kp}+\frac{{\Pi}_{kp}M_{p}}{\sqrt{24\pi
V(\varphi_{b})}},
\end{equation}
where we have neglected $\Pi_{b}$ since it is much smaller
than $\Pi_{kp}$.

The boundary between regions $\mathcal{P}$ and
$\mathcal{K}$ is given by the equality of kinetic and
potential energy
\begin{equation} \label{piatkp}
\Pi_{kp}=\pm\sqrt{2V(\varphi_{kp})} .
\end{equation}
The sign in the above equation depends on whether one is in
the region of increasing or decreasing $\varphi$. For
concreteness, we will consider trajectories with $\Pi > 0$.
Substituting $\Pi_{kp}$ from above equation into
(\ref{PSboundary1}) and keeping in mind the approximation of V
being constant will lead to:
\begin{equation} \label{PSboundary2}
\varphi_{b}=\varphi_{kp}+\frac{M_{p}}{\sqrt{12\pi}},
\end{equation}
Solving (\ref{region2eq2}) for $\Pi$ we can also write the
equation for the trajectories in phase space while they are in
${\mathcal P}$,
\begin{equation} \label{motionP}
\Pi(\varphi)=\Pi_{kp}\Biggl(1 +\frac{ (\varphi_{kp}-\varphi)
\sqrt{12 \pi}}{M_{p}} \Biggr).
\end{equation}

We now want to extend the trajectories back further to determine
which of the trajectories which begin in region $\mathcal{K}$ 
will lead to successful
inflation, where $\mathcal{K}$ is the region of kinetic energy
domination, i.e. $\frac{1}{2} \Pi^{2} > V(\varphi)$. Here we are
interested in trajectories that will loose enough kinetic energy
(due to the Hubble friction) to eventually reach region
$\mathcal{P}$. In region $\mathcal{K}$ we can ignore the effects
of the potential and (\ref{eom1}) and (\ref{eom2}) become:
\begin{eqnarray} \label{region1eom1a}
\dot{\Pi}+3H\Pi=0\\
\label{region1eom2a} H^{2}=\frac{4\pi}{3 M^{2}_{p}}\Pi^{2}.
\end{eqnarray}
These can be integrated and we find:
\begin{equation}\label{region1eom2b}
\varphi={\varphi_{kp}}+\frac{M_{p}}{\sqrt{12\pi}}\ln \Bigl(
\frac{{\Pi}_{kp}}{\Pi} \Bigr),
\end{equation}
where - as used earlier - $\varphi_{kp}$ and ${\Pi}_{kp}$ are the 
values at the boundary between regions $\mathcal{K}$ and $\mathcal{P}$.  
These regions meet along the curve $\frac{1}{2}\Pi^{2}=V(\varphi)$. We
can solve (\ref{region1eom2b}) for $\Pi$ to find the equation for
trajectories in $\mathcal{K}$:
\begin{equation} \label{k1}
\Pi(\varphi)=\Pi_{kp} \exp \Bigl[ - \frac{\sqrt{12
\pi}}{M_{p}}(\varphi-\varphi_{kp})\Bigr]
\end{equation}
These trajectories continue all the way until $\frac{1}{2}
\Pi^{2} > V(\varphi)$ breaks down or they reach the boundary
of phase space as given by the particular theory (e.g. Planck 
energy density in the case of chaotic inflation, or the limiting
value of $\varphi$ if the inflaton corresponds to the separation
of two branes in an extra dimension and the radius of this extra-dimensional
space is bounded).

Now that we have developed the necessary equations to describe the
full phase space, let us consider the cases of small and large
field inflation separately.

\subsection{Small Field Inflation}

Let us first focus on the case of small field inflation models
such as new inflation. An example of a potential which leads
to small field inflation is sketched in Figure 1 (lower panel). In this
case, during the period of slow roll inflation $\varphi$ is close
to $\varphi = 0$. 

We want to calculate the whole volume of phase 
space that leads to successful slow roll inflation. Most of this phase
space consists of points in ${\mathcal P}$ and ${\mathcal K}$ whose
trajectories enter the region ${\mathcal S}$. In particular,
to obtain enough e-foldings of slow roll inflation, the
value of $\varphi_b$ for such a trajectory must obey
\begin{equation} \label{smallcond}
|\varphi_b| \, \leq \, |\varphi_b^{60}| \, .
\end{equation}
Our method will be to follow the two limiting phase space trajectories
(those for which the inequality in (\ref{smallcond}) is
saturated and which are labelled $\gamma^l$ and $\gamma^u$ in
Figure 1) back into regions ${\mathcal P}$ and ${\mathcal K}$
and then to add up the corresponding phase space volumes between the
two curves in both regions. Note that since the volume of ${\mathcal S}$
is negligible in comparison to the volume of ${\mathcal P}$ (in
particular, the value of $\Pi$ is negligible in region ${\mathcal S}$
compared to $\Pi_{pk}$), we can neglect the contribution of the small
subset of phase space within ${\mathcal S}$ that leads to enough 
e-folding of inflaton.

The calculation involves extending the trajectories through the region
$\mathcal{P}$ by making use of (\ref{motionP}).
Making use of (\ref{PSboundary2}) and assuming that
the condition (\ref{condforV}) is not violated, we obtain the following  
range of values of $\varphi_{kp}$ which lead to trajectories
with successful slow roll inflation:
\begin{equation} \label{inequal1}
\Biggl|\varphi_{kp}+ \frac{M_{p}}{\sqrt{12 \pi}} \Biggr| \leq
\varphi_{b}^{60} \, .
\end{equation}
{}From this result, it is obvious that our approximation of assuming
that the condition (\ref{condforV}) applies throughout region ${\mathcal P}$
is not satisfied in many examples motivated by conventional quantum
field theory in four dimensional space-time, since in those models
we could expect the minimum of the potential to be at a value of
$|\varphi|$ smaller than the value of $|\varphi_{kp}|$ resulting
from (\ref{inequal1}).

According to Equation (\ref{motionP}), trajectories in ${\cal P}$ are
parallel straight lines with a negative slope which extend all the way to
$\Pi_{kp}$, and the above equation determines the side borders for
trajectories leading to successful slow roll inflation. 

Combining these results, we find that the volume $\rho_{p}$ of phase space
contained in region ${\mathcal P}$ leading to successful slow roll inflation 
is
\begin{eqnarray}\label{rhop}
\rho_{p}&=& 2\varphi_{b} | \Pi_{kp} |\\ \nonumber
        &=&2\varphi_{b}^{60}\sqrt{2V(\varphi_{b}^{60})} \, .
\end{eqnarray}

{}For a general potential ${V}$ for which the condition (\ref{condforV}) is
not satisfied throughout,  we have to find the equations of the 
trajectories using
equation (\ref{generalp}), and use the results to obtain the phase space 
points $(\varphi^{l}_{kp},\Pi^{l}_{kp})$ and
$(\varphi^{u}_{kp},\Pi^{u}_{kp})$ which form the end points of the 
boundary between ${\mathcal P}$ and ${\mathcal K}$ in the subspace
of phase space leading to successful slow roll trajectories. Here,
the superscripts $l$ and $u$ stand for the lower and the upper
boundary trajectories, respectively. In this case,
the volume $\rho_{p}$ is:
\begin{equation}
\rho_{p}=\Biggr|\int_{\Pi_{b}}^{\Pi^{u}_{kp}} \varphi^{u}_{p}(\Pi)
\; d\Pi -\int_{\Pi_{b}}^{\Pi^{l}_{kp}} \varphi^{l}_{p}(\Pi) \;
d\Pi,\Biggr| \, .
\end{equation}

{}Finally, we must determine the phase space volume $\rho_k$ within region 
$\mathcal{K}$ bounded by the two trajectories $\gamma^u$ and
$\gamma^l$, trajectories which in region ${\mathcal K}$ obey
the equation (\ref{k1}). However, this time there arises the 
crucial issue of how far back to integrate the trajectory.
First, for the classical field theory analysis to be valid, the energy
density cannot exceed the Planck density, and this imposes an
upper limit on the allowed value of $\Pi$, namely
\begin{equation} \label{picut}
| \Pi | \, \leq \, \sqrt{2} m_{pl}^2 \, .
\end{equation}
Next, the domain of $\varphi$ may be bounded according to the
physics which determines what the inflaton is. For example, if the
inflaton corresponds to the separation between two branes in
a transverse compact dimension, then $\varphi$ is bounded by the
radius of the compact dimension. Finally, in  
the case of potentials which rise again as $|\varphi| \rightarrow \infty$,
the condition, $\frac{1}{2} \Pi^{2} > V(\varphi)$ will cease to hold at
sufficiently large values of $|\varphi|$ \footnote{The region of phase
space beyond this bound would again be an inflating region, but not one
corresponding to small field inflation.}.

If there is a cutoff of the type described above, we will denote the
cutoff values of the phase space coordinates by  
$\varphi_{c}$ and $\Pi_c$. 

We obtain the part of the phase space volume $\rho_k$ 
contained within ${\mathcal K}$ corresponding to successful slow roll
inflation by integrating the trajectories $\gamma^l$ and $\gamma^u$
through region ${\mathcal K}$ using Equation (\ref{k1}) from 
$\varphi = \varphi_{kp}$ to $\varphi = \varphi_c$, and computing
the phase space volume between the curves:
\begin{equation}\label{rhok}
\rho_{k}= \biggr|\int_{\varphi^{u}_{kp}}^{\varphi_{c}}
\Pi^{u}(\varphi) d\varphi
-\int_{\varphi^{l}_{kp}}^{\varphi_{c}} \Pi^{l}(\varphi)
d\varphi\biggr|,
\end{equation}
where $\Pi^{u}$ and $\Pi^{l}$ are given by (\ref{k1}) evaluated
at
\begin{equation}\label{phikp}
\varphi_{kp}^{u,l}=-\frac{M_{p}}{\sqrt{12
\pi}}\pm\varphi_{b}^{60} \, .
\end{equation}
Note that there is an equal size region in phase space obtained
by reflection about the origin of phase space. 

Since $\varphi_{b}^{60}$ is very small compared to both $M_p$ and to
the typical energy scale in the potential, we can first of all
replace the lower integration limit in both integrals in (\ref{rhok})
by the average of the two lower limits, which is $-M_p / \sqrt{12 \pi}$.
Next, we can
Taylor expand the difference between $\Pi^{u}$ and $\Pi^{l}$
(making use of (\ref{k1}))
about the value of $\Pi$ obtained from the value of $\varphi$ along
the central of the three diagonal curves in the upper left quadrant
of Figure 1, with the result
\begin{eqnarray}\label{pkupkl}
\Pi^{u}_{k}(\varphi) &-& \Pi^{l}_{k}(\varphi) \\ 
&\simeq& \,
2\frac{\sqrt{12 \pi}}{M_{p}}\varphi_{b}^{60}
\sqrt{2V(\varphi_{b})}\exp\Biggr[-1 -\frac{\sqrt{12
\pi}}{M_{p}}\varphi \Biggr ] \, , \nonumber
\end{eqnarray}
where we have made use of (\ref{piatkp}) to replace the factor $\Pi_{kp}$
appearing in (\ref{k1}) by the term involving the potential.

Making use of this approximation, it is easy to evaluate equation
(\ref{rhok}) and find $\rho_{k}$:
\begin{equation}\label{rhok2}
\rho_{k} \, \simeq \, 2\varphi_{b}^{60}\sqrt{2V(\varphi_{b}^{60})} 
\Biggr[\exp(-1 - \frac{\sqrt{12 \pi}}{M_{p}}\varphi_{c})
- 1\Biggr] \, .
\end{equation}
Recall that $\varphi_{c}$ is negative due to the assumption (\ref{phikp})
on which half of phase space we are considering. Note also
that the second term in the square brackets in (\ref{rhok2})
cancels with the positive contribution of $\rho_p$, thus yielding
\begin{equation} \label{rhok3}
\rho_k + \rho_p \, \simeq \, 2\varphi_{b}^{60}\sqrt{2V(\varphi_{b})}e^{-1}
\exp(-\frac{\sqrt{12 \pi}}{M_{p}}\varphi_{c}) \, .
\end{equation}

Note that in the above, we have assumed that the approximation 
(\ref{condforV})
is valid in the entire region ${\mathcal P}$. For typical
symmetric double well potentials with minima at $|\varphi| = \eta$, this
is only satisfied for values of $\eta$ larger than
$M_p / \sqrt{12 \pi}$ \footnote{If the condition (\ref{condforV})
is not satisfied, our result yields an upper bound on the phase
space of successful slow roll inflation.}. For inflationary models
in the context of grand unified field theories, this is not a realistic
assumption, however in the cases of brane inflation considered in the
following section, the assumption more likely to be justified (as long
as the brane inflation model does not involve a hierarchy of scales of
internal dimensions).

Let us consider two concrete examples. In the first example we assume
that the potential keeps decreasing (towards the value zero) at large
values of $|\varphi|$, and that there is a cutoff on the range of values
of $|\varphi|$ at $|\varphi| = |\varphi_c| = M_p$
\footnote{It is easy to check that in this example the trajectories
stay below the cutoff line (\ref{picut}) if the value of the potential
at the origin is several orders of magnitude lower compared to the
Planck density, as it must if inflation is not to generate a too
large amplitude of gravitational waves.} In this case, we obtain
\begin{equation} \label{psvol1}
\rho_{k} + \rho_{p} \, \simeq \, 
2 |\varphi_b^{60}| \sqrt{2 V(\varphi_b^{60})} exp(\sqrt{12} \pi - 1) \, .
\end{equation}
The total allowed phase space volume $\rho_T$ is
\begin{equation} \label{phst}
\rho_T \, \simeq M_p^3
\end{equation}  
and hence the fraction $f$ of the phase space volume which leads to
successful slow roll inflation is
\begin{equation} \label{res1}
f \, \sim \, {{\varphi_b^{60}} \over {M_p}} 
{{\sqrt{2 V(\varphi_b^{60})}} \over {M_p^2}} c_1 \, ,
\end{equation}
where $c_1$ is the contant appearing in the exponential in (\ref{psvol1}).

If we compare the result (\ref{res1}) with what would be obtained by
considering only the configuration space constraint on the initial
conditions for successful slow roll inflation, we find that the
constraint is more severe by a factor which is given by the second
ratio on the right hand side of (\ref{res1}).

In the second example, we consider a model of the type proposed in
new inflation, with a potential of Coleman-Weinberg \cite{Coleman}
type
\begin{equation}
V(\varphi) \, = \, \lambda \varphi^4 \bigl[{\cal{\ln}}(\varphi^2 / \eta^2)
- 1/2\bigl] + {1 \over 2} \lambda \eta^4 \,
\end{equation}
with $\eta$ chosen such that (\ref{condforV}) is satisfied in the
entire region ${\mathcal P}$.
This potential rises at large values of $|\varphi|$, and thus the
cutoff value $\varphi_c$ is determined either by the potential
energy becoming equal again to the kinetic energy, i.e. 
\begin{equation} \label{phic2}
V(\varphi_{c}) \, = \, V(\varphi_{kp}) \exp \Biggr[
\frac{2\sqrt{12\pi}}{M_{p}}(\varphi_{kp}-\varphi_{c})\Biggr] \, ,
\end{equation}
or by the kinetic energy density reaching the boundary (\ref{picut}),
which will occur at the value $\varphi_{l}$ determined by
\begin{equation} \label{phic1}
{{\sqrt{12 \pi}} \over {M_p}} |\varphi_l| \, = \,
{\cal{\ln}}\bigl({{M_p^2} \over {\lambda^{1/2}\eta^2}} \bigr) \, .
\end{equation}
For small values of the potential at the origin, it is the condition
(\ref{phic1}) which determines the largest value of $|\varphi|$.
Inserting the result into (\ref{rhok3}), we otain
\begin{equation} \label{res2}
\rho_{k} + \rho_{p} \, \simeq \, 2 |\varphi_b^{60}| \sqrt{2 V(\varphi_b^{60})} 
\lambda^{-1/2} \bigl( {{M_p} \over {\eta}} \bigr)^2 \, .
\end{equation}
The total phase space $\rho_T$ is determined by (\ref{picut}) and by
the largest value of $|\varphi|$ for which $V(\varphi) < M_p^4$:
\begin{equation} 
\rho_T \, \sim \, \lambda^{-1/4} M_p^3 \, .
\end{equation}
Hence, the fraction $f$ of phase space which leads to
successful slow roll inflation is given by
\begin{equation}
f \, \simeq \, 2{{|\varphi_b^{60}|} \over {\lambda^{-1/4} M_p}} \, ,
\end{equation}
which is the same order of magnitude as the fraction of configuration space
which leads to successful slow roll inflation. Thus, in this example
including nonvanishing initial momentum does not lead to a reduction
in the relative phase space of initial conditions yielding slow roll
inflation compared to the fraction of configuration space yielding
slow roll inflation assuming vanishing initial momenta. The reason
is that for any initial field value, we can choose a finely tuned
initial momentum for which the field will roll up the potential and
land near the origin with vanishing momentum.

\subsection{Large Field Inflation}

The fraction of the energetically allowed phase space of initial 
conditions which leads to successful slow roll inflation is of
order one in most large field inflation models. This is not hard
to understand: First of all,
the volume of configuration space which is energetically
allowed and also gives sufficient inflation is of order unity. Second,
the phase space trajectories approach the slow rolling
curve $\Gamma$ sufficiently fast such that taking the
freedom in the choice of the initial momentum into account
does not lead to any restriction on the allowed phase space which
gives slow roll inflation.

To be specific,
let us consider the chaotic inflation model given by the potential
\begin{equation}
V(\varphi) \, = \, {1 \over 2} m^2 \varphi^2 \, ,
\end{equation}
where the constraints on the amplitude of the fluctuations spectrum
limits the mass to be $m / M_p < 10^{-6}$.

In this example, the slow roll trajectory $\Gamma$ is given by
\begin{equation}
\Pi \, = \, (12 \pi)^{-1/2} m M_p \, .
\end{equation}
The boundary between regions ${\mathcal P}$ and ${\mathcal K}$ is given
by
\begin{equation}
\Pi \, = \, m \varphi \, ,
\end{equation}
and thus the slow roll trajectory ends at the value
\begin{equation}
\varphi_b \, = \, (12 \pi)^{-1/2} M_p \, .
\end{equation}
The value $\varphi_b^{60}$ ($|\varphi| > |\varphi_b^{60}|$ required in
order to have sufficient slow roll inflation) is slightly larger than
the above value, about $|\varphi_b^{60}| = 3 M_p$.

In this example, we can solve for the trajectories in Region ${\mathcal P}$
without making the assumption that the potential is approximately
constant. The result is
\begin{equation} \label{traject}
\Pi(\varphi) - \Pi(\varphi_b) \, = \,
(3 \pi)^{1/2} {m \over {M_p}} \bigl( \varphi^2 - \varphi_b^2 \bigr) \, .
\end{equation}
To get a lower bound on the phase space of initial conditions which
yield sufficient slow roll inflation we will extend the above
trajectories (\ref{traject}) beyond the boundary between ${\mathcal P}$
and ${\mathcal K}$. Taking $M_p$ to be negligible compared to $|\varphi|$
and neglecting $\Pi(\varphi_b)$ compared to $\Pi(\varphi)$, we find that the
trajectories (\ref{traject}) hit the phase space boundary
$\Pi = M_p^2$ at the value $\varphi = \varphi_l$ given by
\begin{equation}
\varphi_l \, = \, - (3 \pi)^{-1/4} M_p \bigl( {{M_p} \over m} \bigr)^{1/2} \, .
\end{equation}
Thus, a lower bound on the allowed phase space of initial conditions which
give sufficient slow roll inflation is obtained by considering the
volume of phase space with $\varphi < \varphi_l$ and arbitrary $\Pi$.
Thus, the fraction $f$ of phase space yielding successful slow roll
inflation is bounded by
\begin{equation}
f \, > \, {{\varphi_c - |\varphi_l|} \over {\varphi_c}} \,
\simeq 1 - (3 \pi)^{-1/4} \bigl( {m \over {M_p}} \bigr)^{1/2} \, ,
\end{equation}
where $\varphi_c$ is the value of $\varphi$ for which the potential
energy reaches Planck density.

\section{Application to Brane Inflation}

In this section we will apply the methods developed in section two
to some current models of brane inflation.  Many models of brane
inflation have been proposed over the years 
\cite{Dvali,Burgess1,Shafi,Burgess2,Juan,Tye,Zavala,Kallosh1,Kallosh2,Blumenhagen}. Let us
briefly review the common features of these models (a more
detailed review can be found in \cite{Quevedo} and references
within).  The starting point is a stack or at least a pair of
parallel p-branes.  The p-branes are taken to be large in three
spatial dimensions and the remaining $(d_{||}=p-3)$ dimensions of
the brane are wrapped on the compactified parallel volume,
$v_{||}$.  The branes are separated by a distance $Y$ in the
remaining $d_{\perp}=(9-p)$ transverse dimensions which are also
taken to be compact with volume $v_{\perp}$. The branes appear as
BPS states protected by supersymmetry (SUSY) and are parallel,
static, and stable. This is realized through the potential for the
branes, which vanishes because the forces from the Neveu-Schwarz (NS) sector,
which contains the dilaton and graviton, are exactly canceled by
the forces due to the Ramond-Ramond (RR) charge. 
The idea behind brane inflation is
to break the SUSY to generate a flat, attractive potential between
the branes.  Once a nonzero potential is generated, the inflaton
is identified with the separation $Y$ of the branes in the
transverse volume $v_{\perp}$.  For large $Y$ the potential has a form
expected from Newtonian gravity, which can be understood 
since the dilaton and graviton
are represented by the exchange of closed strings between the
branes in $v_{\perp}$. The branes inflate as they approach each
other in the transverse dimensions, until they get close enough
for a tachyonic mode to develop.  At this point the open strings
connecting the branes become important and the potential becomes
dominated by the tachyon, bringing about the end of inflation in a
way closely related to hybrid inflation \cite{hybrid}.

The main difference between the various brane inflation models is
the way they break SUSY.  One method is to consider
brane/anti-brane pairs moving on a fixed background.  
In this case, the RR charges of the
branes are opposite and the attractive potential of the
NS sector is not completely cancelled, which leads to the
following potential \cite{Burgess1}: 
\beq \label{potential1}
V(Y)=A-\frac{B}{Y^{d_{\perp}-2}}, 
\eeq
where 
\bea A & \equiv & 2 T_p v_{||}=\frac{2 \alpha e^{\varphi_d}}{(M_s
r_\perp)^{d_{\perp}}} M_s^2
 M_{p}^2,\\
B& \equiv & \frac{\beta  e^{2\varphi_d}}{M_s^8} T_p^2 v_{||}=
\frac{\beta
\alpha^2 e^{\varphi_d} M_{p}^2}{M_s^{2(d_\perp -2)}r_\perp^{d_\perp}}.
\eea
Here, $T_{p}$ is the p-brane tension, $r_{\perp}$ is the
radius of the transverse dimensions (taken to be constant),
$\beta=\pi^{-d_{\perp}/2} \Gamma\left(\frac{d_{\perp} -2}{2}\right)$,
and $M_{s}$ and $M_{p}$ are the string and Planck mass,
respectively. The dilaton $\varphi_d$ is assumed to be
fixed. The inflaton field $\varphi$ is related to the brane
separation by its canonical normalization, 
\beq \label{canonical}
\varphi \, = \, \sqrt{T_{p}v_{||}\over 2}\;Y=\Bigl( \frac{\alpha
e^{\varphi_d}}{2 (M_{s}r_{\perp})^{d_{\perp}}} \Bigr)^{1/2}
M_{s}M_{p} \; Y. 
\eeq 
When one investigates this potential as a
candidate for the inflaton, it is found by enforcing the
constraints of $\mathcal{N} > {\rm 60}$ e-foldings and the COBE
normalization that one cannot obtain successful inflation because
one of the slow-roll conditions fails. Namely, the slow roll
parameter ${\tilde{\eta}}$ (conventionally denoted by $\eta$)
defined by
\beq
{\tilde{\eta}} \, = \, M_p^2 {{V''} \over V}
\eeq
is given by 
\beq \label{eta1}
{\tilde{\eta}} \sim -\beta (d_{\perp}-1)(d_{\perp}-2) \Bigl(
\frac{r_{\perp}}{Y} \Bigr)^{d_{\perp}}, 
\eeq 
and hence cannot be made to
satisfy ${\tilde{\eta}} \ll 1$, because the interbrane separation must
satisfy $Y \ll r_{\perp}$

However, this can be remedied if one considers the effects of the
compactification on the potential.  At large distances when $Y
\sim r_{\perp}$ the presence of other branes or orientifold planes
must be considered and this leads to an image effect, which
softens the potential.  In these {\em hypercubic
compactifications} one finds that, for small separations $z$ from the
antipodal point, the potential takes the form \cite{Burgess1} 
\beq \label{zpot}
V(z)\,=\,A-{1\over4}\,C\,z^4\,, 
\eeq 
where $A$ is defined as
before and 
\beq
C=M_s^{-8}\,e^{2\varphi_d}\,T^2_p\,v_\parallel\,
r_\perp^{-(2+d_\perp)}.
\eeq
This model of brane inflation comes under
the small field classification.

To consider the effect of a nonzero velocity for the branes we
proceed by applying the methods of the small field section. We
want to evaluate (\ref{rhok3}) for the given potential
(\ref{zpot}).  The value of $\pbs$ was derived in \cite{Burgess1}
and is given by 
\beq
z_{b}^{60} \simeq (0.1) \; r_{\perp}
\eeq
or in terms of the normalized field 
\beq
\pbs \simeq (0.1) \; r_{\perp} \sqrt{T_{p}v_{||}\over 2} \, . 
\eeq
The cutoff for this model is given by
the size of the transverse dimensions, 
\beq
\frac{l_{\perp}}{2}=\pi
r_{\perp} \, ,
\eeq
and therefore 
\beq \label{critphi}
\varphi_{c}= - \pi r_{\perp} \sqrt{T_{p}v_{||}\over 2} \, .
\eeq
Using these values and
(\ref{canonical}) we find the following expression for the
fraction $f$ of the available phase space \footnote{We have
assumed that the approximation (\ref{condforV}) for the potential
is valid. This may, however, not be the case for realistic values
of $r_{\perp}$ (see (\ref{critphi})) which are large compared
to $M_s^{-1}$. However, in this case the conditions for successful
inflation are even more stringent, and our result yields an upper
bound on the fraction of phase space which yields successful slow roll
inflation.}, 
\bea \label{therho} 
f \, = \, {\rho \over{\varphi_c M_{p}^{2}}} &\sim& 
\bigl (\alpha e^{\varphi_d} \bigr)^{1/2}
(M_{s}r_{\perp})^{-\frac{d_{\perp}}{2}} \Bigl(
\frac{M_{s}}{M_{p}} \Bigr) \\
&& \times \exp{\Bigl( \pi^{3/2}\sqrt{6 \alpha
e^{\varphi_d}}}(M_{s}r_{\perp})^{1-\frac{d_{\perp}}{2}} \Bigr) \, . \nonumber 
\eea

{}From this expression we immediately see that, for compactifications
at the string scale with $M_{s} \sim 1 \; TeV$, the initial
conditons for inflation must be extremely
fine-tuned.  However, in hypercubic compactification models (e.g.
\cite{Burgess1}) $M_{s}$ is usually taken to be close to the
Planck scale and $r_{\perp}$ is left as a tunable parameter, the
only requirement being that $r_{\perp}^{-1} \ll M_{s}$ in order
for the effective field theory arguments to hold. As an example,
if we consider four transverse dimensions in the weak coupling
limit (i.e. $\alpha e^{\varphi_d} \sim 1$) and take $M_{s} \sim
10^{-3} M_{p}$ and $(M_{s}r_{\perp}) \sim 10^{4}$ we find 
\footnote{For large values of $M_s r_{\perp}$, the region ${\mathcal K}$
is negligible compared to the region ${\mathcal P}$, and thus the
exponential factor in (\ref{therho}) is not present - this point is
however irrelevant in terms of the numerical values.} 
\beq
{\rho \over{\varphi_c M_{p}^{2}}} \sim 10^{-11} \, .
\eeq
This result indicates that even in this case, the
phase space of initial conditions is highly constrained within
these models. This conclusion is independent of the number of
transverse dimensions and the choice of compactification radius,
as can be seen by considering other values in (\ref{therho}).

Another way to break the SUSY configuration of the branes is to
introduce an angle $\theta$ resulting from the branes wrapping
different cycles in the compact dimensions \cite{Juan}. This is a more general
approach and setting $\theta=\pi$ gives the brane/anti-brane case.
The potential for these models has a form similar to
(\ref{potential1}). For the case $d_{\perp}=4$ it takes the form
\cite{Tye}, 
\beq \label{potential2}
 V(Y)=\frac{1}{4}T_{4}v_{||}\tan^{2}\theta-\frac{M_{s}^{2}
 \sin^{2}\frac{\theta}{2}\tan\frac{\theta}{2}}{8 \pi^{3}Y^{2}}
\eeq 
These models have several advantages over the limited case of
brane/anti-brane inflation.  First, as pointed out in \cite{Tye}, for
small angles inflation can occur independently of the
compactification method, i.e. hypercubic compactification is not
essential. This can be seen by the modification to (\ref{eta1})
for branes at angles, 
\beq {\tilde{\eta}} 
\sim \theta^{2} \Bigl(\frac{r_{\perp}}{Y}\Bigr)^{d_{\perp}}. 
\eeq 
Thus, for small enough
$\theta$ the conditions for inflation are satisfied.  These models
reduce some of the fine-tuning associated with the size of the
compactification, however, introducing this angle actually makes
the region of available phase space given by (\ref{therho}) more
constrained.  For small angles, the ratio of available phase space
is reduced by a factor of $\theta$ for the momentum constraint and
the constraint on the field introduces a factor of
$(1-\theta^{2/d_\perp})$.  That is, 
\bea \label{therho2}
 f \, = \, \frac{\rho}{\varphi_{c}M_{p}^{2}} &\sim&
\frac{(1-\theta^{2/d_\perp})\theta}{(M_{s}r_{\perp})^{\frac{d_{\perp}}{2}}}
\Bigl( \frac{M_{s}}{M_{p}} \Bigr) \bigl( \alpha e^{\varphi_d} \bigr)^{1/2} \\ 
&& \times \exp{\Bigl( \pi^{3/2}\sqrt{6
\alpha e^{\varphi_d}}}(M_{s}r_{\perp})^{1-\frac{d_{\perp}}{2}}
\Bigr). \nonumber
\eea

In the cases where $Y\sim l_{\bot}/2$, then, as in the case
of the brane-antibrane configuration, the forces by the
images of one brane on the other one should be included. This
implies that the potential (\ref{potential2}) takes a form like 
in (\ref{zpot}),
with $A=\frac{1}{4}T_{4}v_{||}\tan^{2}\theta$ and with $C$ 
proportional to $\sin^{2}\frac{\theta}{2}\tan\frac{\theta}{2}$.
For this potential, the slow roll parameter $\tilde{\eta}$ in 
the case of $d_{\perp}=$ 4 is related to $\theta$ as follows:
\begin{equation}
{\tilde{\eta}} \, \simeq \, 
- 36 \theta^{2}\biggr(\frac{z}{l_{\bot}}\biggr )^{2} \, .
\end{equation}
{}For other values of $d_{\perp}$ similar results can be obtained.
Requiring $|\tilde{\eta}|<\frac{1}{60}$ we get
\begin{equation}
\biggr(\frac{z_{b}}{l_{\perp}}\biggr)\leq \frac {2.1\%}{\theta} \, .
\end{equation}
Once again, we see that by taking small values for $\theta$ 
the fine-tuning associated with configuration space can be reduced.
However, taking into account the
momentum space will cancel this effect since for small values of
$\theta$
\begin{equation}
\sqrt{2V(z_{b})} \, \sim \, \sqrt{\frac{1}{2}T_{4}v_{||}\theta^{2}} \, .
\end{equation}
Inserting this result in (\ref{rhok3}) we obtain,
\begin{eqnarray}
\frac{\rho}{M_{p}^{2}\varphi_{c}} &\sim& 
10^{-2}\biggr(\frac{\alpha
e^{\varphi_d}}{(M_{s}r_{\perp})^{d_{\perp}}}\biggr)^{\frac{1}{2}}(\frac{M_{s}}{M_{p}})\nonumber\\
&\times&\exp \biggr(\pi^{\frac{2}{3}}\sqrt{6\alpha
e^{\varphi_d}}(M_{s}r_{\perp})^{1-\frac{d_{\perp}}{2}} \Bigr) \, .
\end{eqnarray}

These results demonstrate that introducing initial velocities for
the branes can drastically reduce the available phase space for
adequate slow-roll inflation.  This makes the models seem
unnatural from a cosmological standpoint, since one would expect
the early universe to contain a gas of branes in random motion
relative to each other (see e.g. \cite{ABE,Majumdar:2002hy}).  

However, on the string theory side, the
branes are usually taken to be BPS states initially and therefore
are static and have no initial velocities.  Then one gradually
breaks the SUSY, which leads to the models discussed above.
Introducing velocities is also a way to break SUSY and also
requires an additional term to be included in the potential.  For
small velocities this term can be neglected, and we have assumed
in our analysis that this is indeed the case.  From a string
theoretic perspective the question then becomes, why should SUSY
breaking be small?  This remains one of the important unsolved
questions in string theory.

\section{Discussion and Conclusions}

In the context of a cosmological scenario with a hot intial stage,
one cannot assume that initial field momenta vanish. Allowing
for nonvanishing field momenta is known to make the initial condition
problem for certain inflationary models of the {\it small field} type
worse. In this paper, we have studied the constraints on the phase
space of initial conditions for brane inflation models which lead 
to successful slow roll inflation. We have found that in
certain models, in particular in models with branes at an angle,
allowing for nonvanishing initial momenta for the inflaton field
(i.e. for the brane separation) dramatically reduces the phase
space of initial conditions for successful slow roll inflation.
We have traced the reason for this sensitivity to initial momenta
to the fact that these models are closer to the class of {\it small field}
inflationary models than {\it large field} models.

In our analysis, we have neglected the fact that the potential for
the modulus field $\varphi$ is velocity-dependent. Given that our
goal is to derive an upper bound on the phase space of initial
conditions which can lead to successful slow roll inflation, our
approximation appears justified, since velocity dependent terms will
steepen the potential (since they lead to a greater departure from the
BPS condition) and thus make it harder to achieve inflation.

Certain of the proposed brane inflation models avoid the problem
discussed in this paper,
e.g. the model of \cite{Tye} which in field theory language appears
more like a {\it large field} model. However, the solution of the
initial condition problem in this model comes at the cost of
introducing a hierarchy in the scales of the extra dimensions,
a hierarchy which also should be explained in the context of
cosmology. Another brane inflation model which can lead to a large field
inflation scenario (and is thus insensitive to allowing nonvanishing
initial momenta) is the one proposed in \cite{Burgess2} in which the
branes are stuck at orbifold fixed planes but the radius of the
extra dimensions becomes dynamical. 
Note that the initial condition problem is
absent if inflation is topological in nature, i.e. occurs inside
of a topological defect (see \cite{Alexander2,ABR} 
for implementations in the context of brane world scenarios).

If one relaxes the assumption of a hot beginning, and instead
assumes that the initial state only differs slightly from a cold BPS
state, then the initial condition problem discussed in this paper
disappears. However, this requires a significant change in our
current view of initial conditions of the early Universe at the
time when a description in terms of classical general relativity
becomes applicable. It is possible, however, that such initial
conditions may arise from considerations of quantum cosmology.

{\bf Acknowledgments}

This work was supported in part (at Brown) by the U.S. Department of 
Energy under Contract DE-FG02-91ER40688, TASK A. SW was supported
in part by the NASA Graduate Student Research Program.

\end{document}